\documentstyle[12pt]{article}
\newcommand{\be}{\begin{equation}}
\newcommand{\ee}{\end{equation}}
\newcommand{\ea}{\end{array}}
\newcommand{\beqa}{\begin{eqnarray}}
\newcommand{\eeqa}{\end{eqnarray}}
\def\CDalign#1{\bgroup\vcenter\bgroup\tabskip 2pt
       \baselineskip 14pt \lineskip 3pt \lineskiplimit 3pt
       \halign\bgroup &\hfill$##$\hfill\crcr
       #1\crcr\egroup\egroup\egroup}

\newcommand{\gapproxeq}{\lower .7ex\hbox{$\;\stackrel{\textstyle
>}{\sim}\;$}}
\newcommand{\lapproxeq}{\lower .7ex\hbox{$\;\stackrel{\textstyle
<}{\sim}\;$}}

\newcounter{appendice}

\def\thebibliography#1{{\bf REFERENCES\markboth
 {REFERENCES}{REFERENCES}}\list
 {[\arabic{enumi}]}{\settowidth\labelwidth{[#1]}\leftmargin\labelwidth
 \advance\leftmargin\labelsep
 \usecounter{enumi}}
 \def\newblock{\hskip .11em plus .33em minus -.07em}
 \sloppy
 \sfcode`\.=1000\relax}

\begin{document}
\begin{titlepage}
\title{
{\small\hfill FFIA-UV/04-05}\\
{\bf  Supertubes versus superconducting tubes }
\author{
Rub\'en  Cordero$^a$ \footnote{Email:
cordero@esfm.ipn.mx} \,
\,and Efra\'\i n Rojas$^b$ \footnote{Email:
efrojas@uv.mx} \\
{\small\it $^a$ Departamento de F\'\i sica,
Escuela Superior de F\'\i sica
y Matem\'aticas del I.P.N.}\\
{\small\it Unidad Adolfo L\'opez Mateos, Edificio 9,
07738 M\'exico, D.F., MEXICO}\\
{\small\it $^b$ Facultad de F\'\i sica e Inteligencia Artificial, Universidad
Veracruzana} \\
{\small\it Sebasti\'an Camacho 5, Xalapa, Veracruz;
91000, MEXICO}
} }
\maketitle
\begin{abstract}
In this paper we show the relationship between cylindrical
D2-branes  and cylindrical superconducting membranes described
by a generic effective action at the bosonic level.
In the first case the extended objects considered, arose as
blown up type IIA superstrings to D2-branes, named supertubes.
In the second one, the cosmological objects arose from some
sort of field theories.
The Dirac-Born-Infeld action describing supertubes is shown
to be equivalent to the generic effective action describing
superconducting membranes via a special transformation.
\end{abstract}
\begin{center}
{\it (Dedicated to Prof. Alberto Garc\'{\i}a on the ocassion of
his 60th birthday)}
\end{center}
\end{titlepage}

\section{Introduction}

Nowadays, a number of outstanding problems in physics
are currently being worked out using techniques involving
the string theory machinery. Every time new ideas transform
our understanding of string theory. Since their appeareance,
the so-called $Dp$-branes \cite{Gibb,Polch} paved the way for
a fierce study in string theory because is assumed that they
are source for the Ramond-Ramond fields in Type II
theories besides of be a useful tool to explain several
supersymmetric and non-supersymmetric field theories.
Among main features of $Dp$-branes, resides that
10-dimensional superstrings can end on them.
The incorporation of $Dp$-branes in superstring
picture gives a lot of issues in the theory
of solitonic states in nonperturbative string
theory and permits to reveal different aspects of
string/$M$-theory dualities.

Recently, special $D2$-branes named supertubes, have
emerged as worldvolume realizations of some kind
of sigma models \cite{Mateos1,Mateos2},
exhibiting more features for the intringuing $M$-theory.
The role are playing Born-Infeld fields describing
cylindrical $D2$-branes
is to maintain stability against tension, including
supergravity forces. On other hand, at cosmological
context, some field theories predict the existence
of topological defects like strings or domain walls
holding ability of carry some kind of charge
\cite{Vilenkin, Lazarides,Peter,supercond, Larsen, Carter}.
For these superconducting strings/membranes, a scalar field
living onto the worldsheet plays a similar role as before,
i.e., maintain stability against tension. Due to previous
similarities, it is to be expected similar descriptions
in their dynamics, i.e., must exists a transformation
between both dynamical descriptions.

In this paper we shall concentrate on the dynamical similarities
between supertubes and superconducting cylindrical branes
which are extended objects arose at first instants of the
universe. The paper is organized as follows.
In Sect. 2 we explain the notation we use through the paper and
obtain the canonical momenta associated to the embedding variables
by means of variational techniques and with the help of
Noether theorem \cite{CapoGuven, Noether},
in order to identify the energy density. In Sect. 3
we describe the main results about supertubes developed in
\cite {Mateos1}. In Sect. 4 we write the generic effective action
which provide us the dynamics of superconducting extended
objects. The comparison of both dynamics is given in Sect. 5.
Finally we end the paper with some comments.

\section{Noether currents and momenta}
\setcounter{equation}{0}

We consider a D$p$-brane of dimension $d$ evolving
in a $N+1$ dimensional background spacetime with metric
$g_{\mu \nu}$, $\mu,\nu = 0,1\ldots ,N$. The D$p$-brane
worldvolume is an oriented timelike manifold of
dimension $d$, usually denoted by $m$, endowed with an
induced metric $\gamma_{ab}$ from the bulk. If $x^{\mu}=
X^{\mu}(\xi^{a})$
are the embedding functions of the D$p$-brane,
the induced metric on the worldvolume is given by
$\gamma_{ab}=g_{\mu\nu}X^{\mu}{}_{,\,a}X^{\nu}
{}_{,\,b}$.  \,$\xi^{a}$ denotes the worldvolume coordinates.

We demand that the Dirac-Born-Infeld action (DBI) specify the
motion of the $Dp$-brane,
\begin{equation}
S_{BI}=  \alpha \int_m d^{d+1}\xi \,\sqrt{-{\mbox{det}}
(\gamma_{ab} + F_{ab})}\,,
\label{eq:DBIaction}
\end{equation}
where $\alpha$ is the tension of the D$p$-brane,
$F_{ab}= 2\partial_{[a}A_{b]}$
is the electromagnetic tensor associated to the
worldvolume $U(1)$ gauge field $A_a$;
 $a,b=0,1,\ldots,d$.

A general variation of the action (\ref{eq:DBIaction})
always cast out  in the form \cite{CapoGuven, Noether}
\beqa
\delta S_{BI}&=& \int_{m} \,\,\left\lbrace {\cal E}_\mu\,
\delta X^{\mu} + {\cal E}^{a}\delta A_a
+ \sqrt{-\gamma}\,\nabla_a F^{a}[\delta X;\delta A]\right\rbrace ,
\nonumber \\
&=& \int_{m} \,\,\left\lbrace {\cal E}_\mu\, \delta X^{\mu}
+ {\cal E}^a \delta A_a \right\rbrace
+
\int_{\partial m}\left\lbrace  P_\mu \delta X^\mu +
\pi^a \delta A_a \right\rbrace ,
\eeqa
where ${\cal E}_\mu$ and ${\cal E}^{a}$ are the Euler
Lagrange derivatives of ${\cal L}_{BI}$ with respect
to $X^{\mu}$ and $A_a$, respectively; $F^{a}$ is an
operator defined on the worldvolume and it is related
to the momenta associated to the configuration space.
The argument of the $F^{a}$, is indicated within the
square bracket.
$P_\mu$ and $\pi^a$ are canonical momenta associated
to the configuration space. We are interested
in the momentum $P_\mu$ given by
\be
P_\mu = \sqrt{h} T^{ab} \eta_a X_{\mu\,,b}\,,
\label{eq:p}
\ee
where $T^{ab}$ is the energy-momentum tensor
(see (\ref{eq:something})) and $\eta_a$ is the unit
timelike normal vector
on the boundary $\partial m$. In fact, when one perform
a foliation of the worldvolume in spacelike hypersurfaces $\Sigma$,
the corresponding normal vector is $\eta^{a}$.

When the classical equations of motion are satisfied,
${\cal E}_\mu=0$ and ${\cal E}^{a}=0$, these equations
are equivalent to the set
\beqa
T^{ab}K_{ab} ^{i}&=&0\,,
\label{eq:T1}
\\
\nabla_a T^{ab}&=&0\,,
\label{eq:T3}
\\
\nabla_a {\cal J}^{ab}&=&0\,,
\label{eq:T2}
\eeqa
where $T^{ab}$ denotes the symmetric energy-momentum
tensor, ${\cal J}^{ab}$ denotes an antisymmetric
bicurrent density and $K_{ab} ^{i}$
denotes the extrinsic curvature of the
worldvolume, where $i$ runs from $1$ to $N+1-d$, {\it i.e.,}
$i$ labels the number of unit normals to the worldvolume.
Explicitly, $T^{ab}$ and ${\cal J}^{ab}$ are given by
\beqa
T^{ab}&=&\frac{\sqrt{-M}}{\sqrt{-\gamma}}
(M^{-1})^{(ab)}\,,
\label{eq:something}
\\
{\cal J}^{ab}&=& \alpha \sqrt{- M}\,(M^{-1})^{[ab]}\,,
\eeqa
where $(M^{-1})^{ab}$ means the inverse matrix of
$M_{ab}:=\gamma_{ab} + F_{ab}$ and $M = {\mbox {det}} (M_{ab})$.

In order to illustrate the previous results, from
(\ref{eq:DBIaction}), the corresponding Lagrangian density
for $D2$-branes is
\be {\cal L}=
\sqrt{-\gamma}\left( 1 + \frac{1}{2}F_{ab}F^{ab} \right) ^{1/2}\,,
\ee
where $\gamma$ denotes the determinant of $\gamma_{ab}$. In such case,
we are able to write explicitly both the symmetric and antisymmetric
parts of the inverse matrix of $M_{ab}$, namely
\beqa
\left( M^{-1}\right)^{(ab)}&=& \gamma^{ab} -
\left( 1+ F^2  \right)^{-1/2} \,F^{c(a}F_c{}^{b)}\,,\\
\left( M^{-1}\right)^{[ab]}&=& - \left( 1+ F^2  \right)^{-1/2}\,
F^{ab}\,,
\eeqa
where we use the notation $F^2 = (1/2)F_{ab}F^{ab}$.

\section{Supertubes}

Brane expansion is an interesting effect in physics.
A current topic in $Dp$-brane physics concerns expansions
from one configuration to another. Interaction of
$Dp$-branes with external fields (not necessarily at
supersymmetric level), under certain circumstances,
produce expansions in order to stabilize a given
configuration. It has been recently shown \cite{Mateos1}
that type IIA strings can blow up to produce cylindrical
$D2$-branes, being the angular momentum the stabilizer
for maintaining a finite radius for the
cylinder.

Following \cite{Mateos1}, consider a static tubular $D2$-brane with
radius $R(z,\phi)$, embbeded in a $N=10$ Minkowski spacetime \be
ds^2 = -dT^2 + dZ^2 +dR^2 + R^2 d\Phi^2 + ds^2(E^6), \ee with
worldvolumen coordinates $T=t, Z=z, \Phi=\phi$, where the axis is
along the $z$ direction. The induced metric is given by \be ds^{2} =
-dt^{2} + dz^{2} + R^{2}d\phi^{2} + (\partial_z R \,d{z} +
\partial_\phi R \,d\phi)^{2}\,. \ee Assuming time independent
Born-Infeld fields via the 2-form field strength \be F = E dt \wedge
dz + B dz \wedge d\phi\,, \ee the $D2$-brane Born-Infeld Lagrangian
(\ref{eq:DBIaction}) is \be {\cal L} = \sqrt{(R^{2} + R_\phi
^{2})(1-E^{2}) + B^{2} + R^{2}R_z ^{2}} \,, \label{eq:lag} \ee where
$R_\phi = \partial_\phi R$ and $R_z = \partial_z R$. Defining the
electric displacement $\Pi \equiv \partial {\cal L}/\partial E$, the
associated Hamiltonian density to (\ref{eq:lag}) results \be {\cal
H} = \Pi \,E - {\cal L}\,. \ee There is a relation between the
electric field and the electric displacement given by \be E =
\frac{\Pi}{R}\sqrt{\frac{B^{2}+R^{2}}{\Pi^{2}+R^{2}}}\,, \ee such in
the case when $R$ is constant and the energy density becomes \be
{\cal E} = {1\over R} \,\sqrt{(\Pi^{2} + R^{2})(B^{2} + R^{2})}\,.
\label{eq:energy} \ee The $z$ independence of the electric
displacement comes from the fact that of $\Pi$ is subject to the
Gauss law constraint.

\section{Superconducting tubes}

To describe superconducting branes we consider
the effective action \cite{Vilenkin, Lazarides, Peter,
supercond,Larsen,Carter}
\be
S_c = \int_m d^{d+1}\xi \,\sqrt{-\gamma}\,L(\omega)\,,
\label{eq:super}
\ee
where $L(\omega)$ denotes a specific model depending
on internal fields acting on the brane through the
combination $\omega := \gamma^{ab}\phi_{,\,a}
\phi_{,\,b}$. There is an important model developed
by Nielsen \cite{Nielsen}, which consists basically
in consider a scalar field living on the worldvolume
which, under certain circumstances, prevents the collapse.
For the case of a superconducting membrane, this model
is given by
\be
L = \sqrt{k_1 + k_2\omega}\,,
\label{eq:lagrangian}
\ee
where $k_1$ and $k_2$ are constants. The corresponding
equations of motion are (compare with (\ref{eq:T1})
and (\ref{eq:T2}))
\beqa
T^{ab}K_{ab} ^{i} &=&0\,,
\label{eq:motion}
\\
\nabla_a J^{a} &=&0
\label{eq:consj}
\,,
\eeqa
where the explicit form of the energy-momentum
tensor is
\be
T^{ab} = L (\omega)\gamma^{ab} -2\frac{dL}{d\omega}
\gamma^{ac}\gamma^{bd}\phi_{,\,c}\phi_{,\,d}\,.
\ee
Note that the second equation (\ref{eq:consj})
corresponds to a conservation
law being the conserved current defined as $J^{a}= 2\,
\frac{d L}{d \omega}\,\gamma^{ab}\,\nabla_b \phi$.

We assume a background 4-dimensional Minkowski metric of the form
\be
ds^2 = - dt^2 + dr^2 + r^2d\theta^2 + dz^2\,,
\ee
and we specialize to a superconducting tube described by the embedding
\be
x^\mu (t,
\theta, z) = \left(
\begin{array}{c}
t\\
r(t)\\
\theta \\
z
\end{array}
\right) \,,
\ee
where the radius $r=r(t)$ is $\theta$ and $z$
independent. The induced metric on the tube is given by
\be
ds^2 = (-1 + \dot{r}^2)\,dt^2 + r^2\,d\theta^2 + dz^2 \,.
\ee

Under the following assumption for the scalar field, $\phi = \phi_1
(t) + N\,\theta$, and from the conservation Eq. (\ref{eq:consj}) we
get how evolve in time the scalar field \be \dot \phi = -
\frac{\Omega \sqrt{1-\dot{r}^2}}{2r(dL/d\omega)}\,, \ee where
$\Omega$ is an integration constant and $N$ is a constant to be
determined. It is straightforward to obtain the form of $\omega$ \be
\omega = \frac{N^2 - (\Omega^2k_1/k_2) ^2}{r^2 + (\Omega^2/k_2)}\,.
\ee Handling the previous results allow us to write the explicit
dependence of the Lagrangian (\ref{eq:lagrangian}), \be L=
\sqrt{\frac{{k_1 r^2 + k_2 N^2}}{r^2 + (\Omega^2/k_2)}}\,. \ee

\section{Mechanical equivalence between tubes}

Now we turn to show the similarity in dynamics between
supertubes and superconducting tubes.
When $R$ is $\theta$ and $\phi$ independent, this fact
allows us to rewrite the Lagrangian as
\be
{\cal L} = - \sqrt{R^{2}(1 - E^{2}) + B^{2}}\,.
\ee

The constant $N$ appearing in the definition of the scalar field,
can be computed in terms of $\Pi$ using the relation between
(\ref{eq:consj}) and (\ref{eq:T3}). Inserting the results for
$L(\omega)$ and $dL/d\omega$ in the definition of the electric displacement,
we get $\Pi = (k_2/k_1)^{1/2}\,N$.

We look at the zero-zero component in the
energy-momentum tensor
\beqa
T^{00} &=& L(\omega)\gamma^{00} - 2\,\frac{d L}{d \omega}
(\gamma^{00})^{2} \dot{\phi}^{2} \nonumber \\
&=& \frac{1}{r^{2}(-1 + \dot{r}^{2})}\,\sqrt{(k_1 r^{2} + k_2
N^{2})(k_2 r^{2} + \Omega^{2})}
\eeqa
such that, for the static
configuration $r=R$ the energy density can be extracted from the
time component of the momentum vector (\ref{eq:p}), i.e.,
\beqa
{\cal E} &=& {1\over R}\,\sqrt{(k_1 R^{2} + k_2 N^{2})(k_2 R^{2} +
\Omega^{2})} \,,\nonumber
\\
&=&{1\over R}\,\sqrt{k_1 (R^{2} + \pi^{2})(k_2 R^{2} +
B^{2})} \,.
\eeqa
If we choose $B=\Omega$ and $k_1 = k_2 =1$
then $\Pi = N$ and we get the energy in accordance
with the result arose from Born-Infeld action
(\ref{eq:energy}), specifically for the case of supertubes
\cite{Mateos1}. It is well known \cite{Vilenkin,
supercond,Carter} that equilibrium configurations for
superconducting extended objects are reached thanks
to the current produced by the scalar field, avoiding
the collapse. Actually, $\phi$ plays a similar role as
the angular momentum for the supertubes case.

\section{Concluding remarks}

 We have shown an equivalence, at the bosonic
level, between supertubes and superconducting tubes. Superconducting
membranes are described uniquely by a scalar field living on its
worldvolume. The energy in both tube cases turns out to be same if
we relate the conserved type IIA string charge and the D0-brane
charge per unit length carried by the tube with the current and the
charge of the superconducting tube, respectively. The next task is
to explore the possibility to introduce supersymmetry in the
superconducting membranes approach and show explicitly its character
1/4 supersymmetric in order to complete the proof of total
equivalence between supertubes and superconducting tubes. The last
issue is under current investigation.

\section{Acknowledgments}

ER and RC received partial support from CONACyT grant CO1-41639.
ER acknowledges also partial support from PROMEP 2003-2004.
RC thanks COFAA and CGPI grant 20030642 as well as SNI-CONACyT
for financial support.

\vfill\break
\bibliographystyle{unsrt}

\end{document}